\documentclass[conference]{IEEEtran}
\IEEEoverridecommandlockouts

\usepackage{cite}
\usepackage{amsmath,amssymb,amsfonts}
\usepackage{graphicx}
\usepackage{textcomp}
\usepackage{xcolor}
\usepackage{tabularx}
\usepackage{makecell}
\def\BibTeX{{\rm B\kern-.05em{\sc i\kern-.025em b}\kern-.08em
    T\kern-.1667em\lower.7ex\hbox{E}\kern-.125emX}}

\usepackage{booktabs}
\usepackage{algpseudocode}
\usepackage{algorithm}
\usepackage{amsmath}
\usepackage{amssymb}
\usepackage{multirow}
\usepackage{subcaption}
\usepackage{hyperref}

\usepackage{fancyhdr}
\usepackage{lipsum}  

\fancypagestyle{firstpage}{
  \fancyhf{} 
  \fancyhead[L]{2024 International Conference on Activity and Behavior Computing} 
  \fancyfoot[C]{\thepage} 
}

\begin{document}
\title{PupilSense: Detection of Depressive Episodes Through Pupillary Response in the Wild

%
}

\author{
\IEEEauthorblockN{
Rahul Islam \href{https://orcid.org/0000-0003-3601-0078}{\includegraphics[scale=0.06]{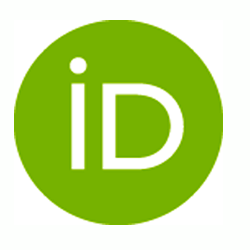}},
Sang Won Bae \href{https://orcid.org/0000-0002-2047-1358}{\includegraphics[scale=0.06]{Figs/orcid.png}}
}
\IEEEauthorblockA{School of Systems and Enterprises\\
Stevens Institute of Technology\\
Hoboken, USA\\
}
}

\maketitle
\thispagestyle{firstpage}  

\begin{abstract}
Early detection of depressive episodes is crucial in managing mental health disorders such as Major Depressive Disorder (MDD) and Bipolar Disorder. However, existing methods often necessitate active participation or are confined to clinical settings. Addressing this gap, we introduce PupilSense, a novel, deep learning-driven mobile system designed to discreetly track pupillary responses as users interact with their smartphones in their daily lives. This study presents a proof-of-concept exploration of PupilSense's capabilities, where we captured real-time pupillary data from users in naturalistic settings. Our findings indicate that PupilSense can effectively and passively monitor indicators of depressive episodes, offering a promising tool for continuous mental health assessment outside laboratory environments. This advancement heralds a significant step in leveraging ubiquitous mobile technology for proactive mental health care, potentially transforming how depressive episodes are detected and managed in everyday contexts.
\end{abstract}


\section{Introduction}
Major Depressive Disorder (MDD) and Bipolar Disorder (BD) collectively impact a significant portion of the global population. Estimates suggest that up to 20\% of individuals may experience one of these conditions over their lifetime \cite{james2018global}. These disorders represent substantial contributors to the global burden of disease, ranking among the most significant causes of disability worldwide \cite{james2018global}. Characterized by recurrent depressive symptoms, they present formidable challenges to mental health and overall well-being. Depression, a multifaceted mental health condition, is marked by a range of symptoms, including persistent sadness, hopelessness, and a loss of interest in daily activities. It often involves low arousal and pleasure, leading to symptoms like feelings of worthlessness, sleep disturbances, and changes in appetite. In severe cases, depression can even lead to thoughts of suicide or self-harm \cite{arnow2014depressive}. The onset and intensity of these symptoms can be influenced by various factors, including hormonal changes, stress levels, environmental triggers, and underlying medical conditions. Given the severity and variability of depressive episodes, understanding their manifestation is crucial for effective intervention. 

Traditional methods for assessing depression typically rely on self-report questionnaires like the Patient Health Questionnaire-9 (PHQ-9) \cite{kroenke2001phq} and Beck's Depression Inventory (BDI) \cite{beck1996beck}. While these tools are widely used and validated, they may be limited by factors such as recall bias, making accurate self-assessment challenging for individuals. In response to these limitations, researchers have explored interventions such as cognitive-behavioral therapy (CBT) \cite{turner2010cognitive}, antidepressant medication \cite{olfson2003relationship}, patient education, lifestyle changes, and social support \cite{hogan2002social}. However, there is a growing recognition of the need for more objective and continuous assessment methods.

This has led to the emergence of smartphone-based passive sensing as a promising approach \cite{chikersal2021detecting, opoku2022mood, pedrelli2020monitoring, wang2014studentlife, saeb2015mobile, mohr2017personal}. These systems leverage the ubiquity of smartphones to passively collect data on social and behavioral signals, offering insights into an individual's depressive state. This method tracks various parameters like GPS data, call/message logs, app usage, sleep patterns, and physical activity, aiming to provide a more holistic and real-time picture of mental health.

Despite its potential, smartphone-based passive sensing faces challenges in practical deployment due to the need for frequent data collection, which can be inconvenient and intrusive. Recent advancements in this field, exemplified by datasets like GLOBEM \cite{xu2022globem, xu2023globem} and frameworks like AWARE \cite{ferreira2015aware}, highlight the ongoing efforts to overcome these challenges and make smartphone-based passive sensing a viable tool for detecting and monitoring depressive symptoms in real-life settings.

Another approach that can be used to assess depressive symptoms is the measurement of physiological changes as this could reveal more objective manifestations of depression which would lead to effective detection of their symptoms.  Previous research has shown that physiological signals, such as electrodermal activity (EDA), external skin temperature, and heart rate, can be reliably assess depression severity with mean absolute error (MAE) ranging between 3.88 and 4.74. The assessment can be facilitated through the use of wearable sensors (such as the Empatica E4) worn on both hands to track physiological data from the user \cite{pedrelli2020monitoring}. 
However, wearable-based physiological sensing approaches can be associated with high costs (approximately \$1,690 per device), which may pose barriers to widespread deployment and could be burdensome for users.
In an effort to monitor objective physiological indicators of depression in clinically depressed patients, researchers have employed a Polar heart strap worn discreetly around the chest beneath clothing, along with skin conductance sensors affixed to the non-dominant hand of the patient \cite{sung2005objective}. While this approach may yield valuable data, the necessity for multiple sensors can impose a burden on the user, and the presence of external hardware on the body may detract from the optimal user experience in daily life.

Given the limitations of current depression detection methods, a fundamental question arises: How can we measure a user's depressive state in a less burdensome, requiring minimal or no effort from the user at a low cost? This paper proposes a solution: assessing depressive episodes by analyzing pupillary responses captured through a burst of photos taken by a smartphone camera in a naturalistic environment. We have developed a deep-learning based pupillary response sensing system called PupilSense and conducted a proof-of-concept study in real-world settings to assess the feasibility of our system.

In this paper, our contributions are twofold:
\begin{itemize}
     \item We have developed an innovative mobile sensing system, deployable in various settings, aimed at helping researchers gather pupillary responses. Our proof-of-concept study revealed a unique potential: this system can assess an individual's depressive state in naturalistic settings, a capability that stands in contrast to many existing tools restricted to data collection within controlled laboratory environments. We conducted a study in naturalistic settings to evaluate the feasibility of PupilSense, achieving an AUROC of 0.71 in detecting depressive episodes of an individual from PHQ-9.

    \item To facilitate researchers in their data collection endeavors, we have made our system's source code and model\footnote{{Our code and model are available at: https://github.com/stevenshci/PupilSense}} accessible to the community. Furthermore, we discuss the implications of using our system, highlighting potential applications and use cases for detecting and addressing mental health issues, including depressive episodes.

\end{itemize}

\section{Background and Related Work}
In this section, we provide a discussion on understanding depression through pupillary response, as well as discuss current advancements in mobile and wearable sensing for depression.

\subsection{Understanding Depression Through Pupillary Response}
Pioneering research in 1996 investigated the relationship between pupillary response and depression by measuring pupillary sensitivity to pilocarpine \cite{sokolski1996increased}. It was discovered that muscarinic sensitivity increases in depression, as evidenced by depressed patients exhibiting significantly greater reductions in pupillary diameter when exposed to increasing concentrations of pilocarpine, a muscarinic receptor agonist, compared to control groups.

A significant body of research on pupillary response in depression revolves around how the pupil responds to emotional stimuli \cite{burkhouse2015pupillary, siegle2001pupillary, sekaninova2019oculometric, silk2007pupillary, steidtmann2010pupil}, and cognitive stimuli \cite{siegle2004pupillary}. Findings indicate that pupil dilation serves as a physiological indicator of emotional and cognitive load in depression. For example, children of depressed mothers with greater dilation to sad stimuli are at higher risk of depression, while depressed adolescents exhibit altered pupil responses to emotional stimuli, indicative of emotional processing impairments. Depressed adults exhibit sustained pupil dilation when processing negative information. However, depressed children show diminished dilation to negative words after initial processing. Pupillometry can be a cost-effective tool in clinical settings to identify at-risk patients. This response to stimuli through pupil dilation varies based on age and specific emotional triggers.

Recently, Schneider et al. \cite{schneider2020pupil} found that individuals displaying severe depressive symptoms exhibit reduced pupil dilation in anticipation of a reward, suggesting a decreased sensitivity to rewards and diminished capacity to experience pleasure. This underscores the potential of using pupil dilation as a marker for depression detection. Despite extensive research on depression detection using physiological data from wearable sensors \cite{pedrelli2020monitoring}, pupillary response during depressive episodes has received limited attention. \textcolor{black}{Previous studies on pupillary response and depression have primarily been conducted in controlled lab settings; however, observing human behavior and physiological responses in natural settings can yield unique insights, as suggested by Picard et al. \cite{picard2016automating}. Studies exploring pupillary responses to everyday stimuli such as smartphone usage, reading, and browsing, have shown promise \cite{miranda2018eye, mirzajani2022dynamic, shen2022pupilrec}, offering an opportunity to examine how real-world interactions affect physiological indicators of depression. Our study aligns with the passive sensing approach advocated by Pedrelli et al. \cite{pedrelli2020monitoring}, Chikersal et al. \cite{chikersal2021detecting}, and Renn et al. \cite{renn2018smartphone}, emphasizing the importance of understanding real-world physiological responses for effective depression interventions.} Therefore, our study aims to model pupillary response during depressive episodes, utilizing opportunistic pupil-iris ratio (PIR) measurement system in a naturalistic setting.

\subsection{Depression Detection Using Mobile and Wearable Devices}
There has been a growing interest in using smartphone and wearable sensor data to detect depression. Chikersal et al. \cite{chikersal2021detecting} monitored behavioral indicators such as Bluetooth usage, phone calls, GPS location, microphone activity, and screen status using the AWARE \cite{ferreira2015aware} system to identify signs of depression in university students, achieving an accuracy of 85.4\%. Asare et al. \cite{opoku2022mood} also used the AWARE platform to monitor behavioral indicators such as sleep patterns, physical movement, mobile phone activity, geographical positioning, and daily emotional assessments based on the circumplex model of affect (CMA) to identify depression, achieving a detection accuracy of 81.43\%. However, these studies focus primarily on the social and behavioral signals of depression, overlooking its physiological aspects. Depression is a multifaceted condition that affects not only social and behavioral but also physiological aspects of an individual, a facet that remains largely unexplored.

While recent studies have used wearable-based physiological sensing to detect depression, as demonstrated by Pedrelli and colleagues \cite{pedrelli2020monitoring}, who utilized data from Empatica wearable devices and smartphones to monitor fluctuations in depression severity, they incorporated various physiological metrics such as electrodermal activity (EDA), external skin temperature, heart rate, and movement data from a 3-axis accelerometer. Additionally, they investigated sleep quality, social engagement, behavioral patterns, and the app usage diversity to elucidate their correlation with depression levels. In a similar vein, Alchieri et al.\cite{alchieriStress} have utilized a multi-sensor approach to discriminate between "stress" and "no stress" states, leveraging physiological data collected from wearable sensors to inform their analysis.  However, such wearable-based physiological sensing solutions come at a high cost (\$1,690 per device) and often require users to wear multiple sensors. In contrast, our approach aims to provide a non-intrusive method for measuring a user's depressive state with minimal or no effort and at little to no cost. We propose a solution for assessing depressive episodes by analyzing pupillary responses captured through a series of photos taken by a smartphone camera during day-to-day activities, in a naturalistic environment.

\section{Pupil Sense}
In this section, we explore the design and development of our Android application, created to measure pupillary response using images taken discreetly by a smartphone's front camera in everyday settings. The application combines theoretical principles with technical expertise for real-world effectiveness. We'll highlight the essential theories behind our design and the technical details that operationalize these concepts, offering an in-depth view of our application's functionality.

\begin{figure*}[h]
    \centering
    \includegraphics[scale=0.5]{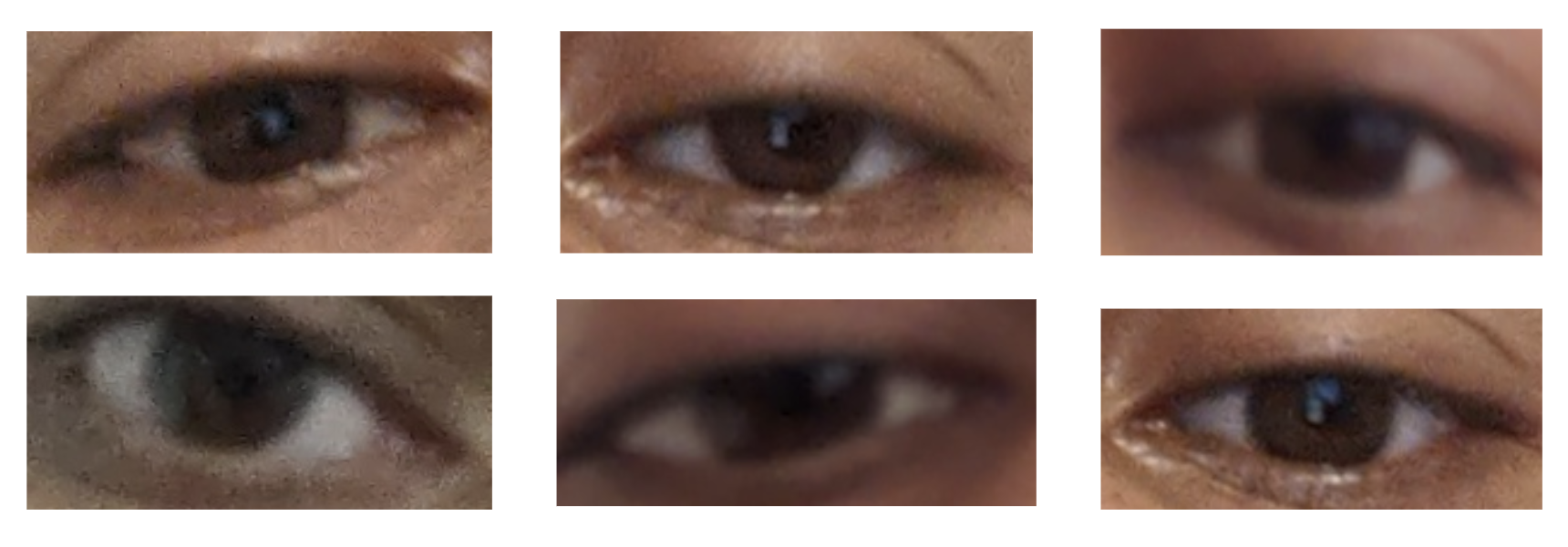}
    \caption{Example of Images in our Dataset for Training Pupi/Iris Segmentation Algorithm collected in Feasibility Evaluation I}
    \label{fig:SampleImagesF1}
\end{figure*}

In pupillometry, researchers use pupil diameter in pixels to measure pupillary response. This method is typically employed when capturing eye images using a pupillometer or eye tracker, where the distance between the camera and the eye remains fixed. In the case of photos taken with smartphones, the scenario differs because users may hold their phones at varying distances. As a result, the apparent size of the pupil can change in images due to the varying distances between the eye and the smartphone camera. In order to address this issue, Tseng et al. \cite{tseng2018alertnessscanner} use pupil-to-iris ratio (PIR) as a measure of pupillary response for detecting alertness in the wild. The authors implement Daugman’s integro-differential operator \cite{daugman1993high} to identify the location of the iris/pupil in the given image. Their method was able to detect pupils with 89.3\% precision in manually collected images by the study participants. However, when they ran a study where the eyes' images were automatically collected, their method performed poorly with only 10.7\% precision in detecting pupils accurately. This may be because when participants manually collect data, the data collection is more controlled and participants are looking at the camera as they are aware of the data collection. While same doesn't hold true for automatically collected eye images where context is uncontrolled. Despite prior works demonstrating high performance in mobile pupilometry measurements during a pupillary light reflex test for at-home scenarios \cite{barry2022home, mariakakis2017pupilscreen, barry2023racially, mcanany2018iphone}, these systems require the user to wear a VR headset-like apparatus. This equipment controls the position of the phone and the lighting that reaches the eyes \cite{mariakakis2017pupilscreen}, making it inefficient for scenarios requiring passive pupilometry. Users may not always have the requisite equipment at hand, yet passive pupilometry is essential because naturalistic observations can yield unique insights into human behavior and the complexity of physiological responses. As noted by Picard et al. \cite{picard2016automating}, everyday environments may unveil behavioral and physiological response patterns that are not observable in controlled settings, such as stressful situations.

Motivated by this we developed a CNN-based algorithm to extract the PIR measure. The method consists of eye image acquisition, eye detection, iris segmentation, and pupil segmentation. We begin by acquiring face images of the user at 2.5hz, the machine learning kit (ML Kit) \cite{mlkit} automatically detects face landmarks. Then, we use the eye landmark points to crop the left and right eye region and send the collected data securely to research for further processing of pupil/iris segmentation described in Section \ref{sec:PIREstimation}. After securely capturing the eye images, original face images are discarded from storage/memory of the phone to perserve privacy. When installed, our app operates as a background service, tracking user activities such as locking/unlocking the phone and usage of apps like WhatsApp and Twitter. These actions trigger a 10-second phase of data collection during which the app employs a photo burst technique, capturing multiple images instead of just one. This approach enhances the likelihood of acquiring higher quality pictures, as detailed in \cite{tseng2018alertnessscanner}. The app captures images at a frequency of 2.5Hz to maintain a balance between image acquisition load and resource usage, ensuring the phone's smooth operation. This approach was based on a study with two participants. Alternatively, a higher frame rate could be achieved by recording as a video stream and analyzing it on a frame-by-frame basis using a codec.

\begin{figure*}[h]
    \centering
    \includegraphics[scale=0.26]{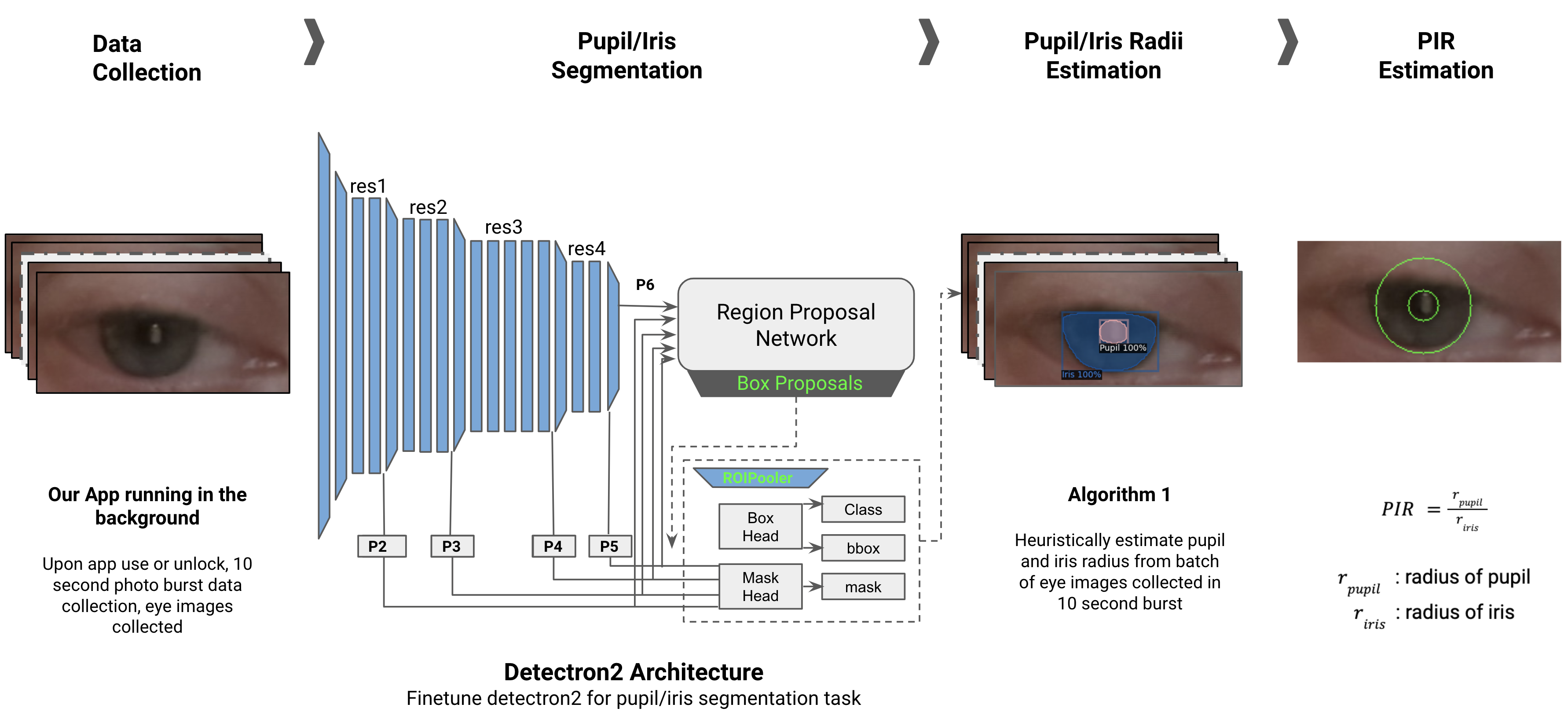}
    \caption{Overview of PupilSense PIR Measurement System}
    \label{fig:PIRMeasurementSystem}
\end{figure*}

\section{Feasibility Evaluation I}
We first conducted a pilot test to obtain the feasibility of estimating pupillary response in the wild. To do this, we enlisted the help of 1 volunteer (T2: tester 2) and the first author (T1: tester 1), who used a Google Pixel 5a and 4 smartphones to gather data for two days. The choice of two participants for feasibility was based on \cite{tseng2018alertnessscanner} a study where they evaluated their system with participants for in-the-wild PIR estimation. Participants carried our app installed in their phones into their daily lives while our app collected data in the background. The volunteers encountered various conditions such as indoor/outdoor, walking/sedentary/changing position, and lighting artifacts such as bright light, daylight/dark from day and night, and some artifacts from ceiling light when data was collected indoors. Both the volunteers were of Asian origin. Figure \ref{fig:SampleImagesF1}, provides samples of images collected during the study. In total, we collected 999 images from T1 and 836 images from T2 from both left and right eye. There were images where the pupil was not visible this includes scenarios such as in dark environments and when images are blurred due to artifacts created from body motion. The results of our app feasibility evaluation were encouraging, and the obtained images were of good quality with few instances where the pupil is not visible, and the eye is closed. Our system acquired eye images without hindering the user device experience at a 10-second photo burst in the background. Section \ref{sec:PIREstimation} explains our process to annotate this collected data and build a method for the task of pupillary response measurement.


\begin{algorithm*}[t]
\begin{algorithmic}[1]
\State $detectron2$ = $Detectron2Model.initialize()$
\Function{estimatePupilIrisRatio}{$**eyeImageSequenceBatch$, $**eyeOpenProbabilities$}
\State $imageSequence\gets \{ \}$\
\For{$image$ $in$ $**eyeImageSequenceBatch$}
    \If{$eye$ $open$ $probability$ $of$ $image$ $>=$ $0.75$}
        \State $append$ $image$ $to$ $imageSequence$
    \EndIf
\EndFor
\State $irisRadii\gets [ ]$\
\State $pupilRadii\gets [ ]$\
\State $eyeCenter\gets [ ]$\
\For{$image$ $in$ $imageSequence$}
    \State $prediction\gets detectron2.predict(image)$\
    \If{$prediction.numClasses()\geq 2$} 
        \If{$prediction.numClasses() > 2$}
            \State $choose$ $Iris$ $and$ $Pupil$ $class$ $that$ $has$  $the$ $maximum$ $prediction$ $score$
        \EndIf
        \State $boundingBox\gets prediction.boundingbox()$\
        \State $irisRadius\gets (boundingBox.iris.bottomRightX() - boundingBox.iris.topLeftX())/2$\
        \State $pupilRadius\gets (boundingBox.pupil.bottomRightX() - boundingBox.pupil.topLeftX())/2$\
        \State $eyeCenterX\gets 
        (boundingBox.pupil.topLeftX() + boundingBox.pupil.bottomRightX())/2$\
        \State $eyeCenterY\gets boundingBox.pupil.bottomRightY() - pupilRadius$\
        \State $append$ $(centerX , centerY)$ $to$ $eyeCenter$\
        \State $append$ $irisRadius$ $to$ $irisRadii$\
        \State $append$ $pupilRadius$ $to$ $pupilRadii$\
    \Else
        \State $skip$ $image$
    \EndIf
\EndFor
\State $irisRadiusFinal\gets irisRadii.mean()$\
\State $pupilRadiusFinal\gets pupilRadii.mean()$\
\State $PIR\gets pupilRadiusFinal/irisRadiusFinal$\
\State $return$ $PIR$ 
\EndFunction
\end{algorithmic}
\caption{\label{alg:estimatePupilIrisRatioMethod} Method for Estimating Pupil Iris Ratio}
\end{algorithm*}

\section{Estimating Pupil-Iris Ratio in the Wild} \label{sec:PIREstimation}
We begin by exploring Daugman’s algorithm used in Tseng et al. \cite{tseng2018alertnessscanner} work in measuring alertness. To our knowledge, it is the only work that performs PIR estimation in the wild. However, their method yields poor performance in our dataset due to different artifacts presented by the environment that the user is in. Their PIR estimation method showed better performance only when eye images were manually collected. For performance in automated eye image collected, they only receive 10.7\% precision in detecting pupils accurately. This is not the case for our method where all the data collection happens automatically without requiring participants to collect manual eye images. Furthermore, the pupil/iris segmentation with Daugman’s algorithm requires 7-10min on average to process one image as recursively try to find the best circle to fit the iris and pupil, rendering it infeasible for real-time processing. To address such challenges,  we then began by exploring different segmentation methods such as SegmentAnything \cite{kirillov2023segment}, and Detectron2 \cite{wu2019detectron2} for our task. These models are not specifically trained for pupil/iris segmentation tasks, so they do poorly in segmenting pupil/iris in eye images. After our initial exploration, we finetuned both SAM and Detectron2 with 15 images and found Detectron2 best suited for our dataset. We use Detectron2 architecture with pre-trained weights to fine-tune the model on our dataset collected in the first feasibility study for the task of pupil/iris segmentation.  Detectron2, an open-source object detection framework, encompasses a multi-component architecture. It begins with a backbone network, often based on ResNet, to extract hierarchical features from input images. These features are then processed through a neck module, typically incorporating Feature Pyramid Networks (FPN), to handle multi-scale object representations. A Region Proposal Network (RPN) proposes candidate bounding box regions, and subsequent Region-of-Interest (RoI) pooling extracts fixed-size feature maps from these regions. Task-specific heads, tailored for various objectives like classification, bounding box regression, and mask segmentation, predict the final outputs. We include the abstract network representation in Figure \ref{fig:PIRMeasurementSystem}. The details of the Detectron2 architecture, please see their paper \cite{wu2019detectron2}.  The full pipeline for estimating PIR from the segmentation obtained from our fine-tuned Detectron2 model is presented in Algorithm \ref{alg:estimatePupilIrisRatioMethod}. The algorithm processes eye image sequences and their associated open probabilities. It begins by filtering out images with low eye-open probabilities ($>0.75$). The Detectron2 model is employed for the retained images to compute pupil and iris segmentation and their bounding boxes. To calculate the PIR we use the two-box method  \cite{martin2015conceptual, mojumder2015pupil, regi2017pupil, sajeevan2017correlation, tseng2018alertnessscanner}, in which two parallel rectangles are drawn along the same axis. The PIR is determined by comparing the width of the pupil bounding box to that of the iris bounding box segmented from Detectron2. This approach has proven effective in accurately estimating the PIR from eye images captured at different angles and distances from the camera. A summary of the method is presented in Fig \ref{fig:PIRMeasurementSystem}.

\subsection{Annotation and Modeling}
To finetune our Detectron2 model, we first annotate the data collected in the first feasibility study. We recruited two research volunteers. One volunteer manually annotates the pupil and iris regions in the eye images. Then, one other volunteer verified the annotation made by the annotator for accuracy. The annotations that don't meet the verification criteria we either discard or send them back to the annotator for correction. In total, we annotated 909 images from both the left and right eye from 2 participants. Figure \ref{fig:SampleImagesF1} shows examples of sample images in our dataset.

Next, we finetune a pre-trained Detectron2 model with weight from \cite{d2goWeights}. We trained the model with the configuration of batch zize=4, learning rate=0.0001 and decay rate=0.1 for 5000 iterations, we kept rest of the configuration as default. To evaluate the performance of the model we use 80/20 split for train and test. Our model achieved 75\% precision, and 78.6\% recall for iris segmentation and 35.1\% precision and 43.4\% recall for pupil segmentation. These evaluation metrics were calculated for IOU .50:.95. Finally, we benchmark the performance and runtime of our method on a new unseen participant (T3: tester 3) for one day of data collected with varying contexts.

\subsection{Feasibility Evaluation II}
\begin{figure}[h]
    \centering
    \begin{minipage}{0.47\textwidth}
        \centering
        \includegraphics[scale=0.3]{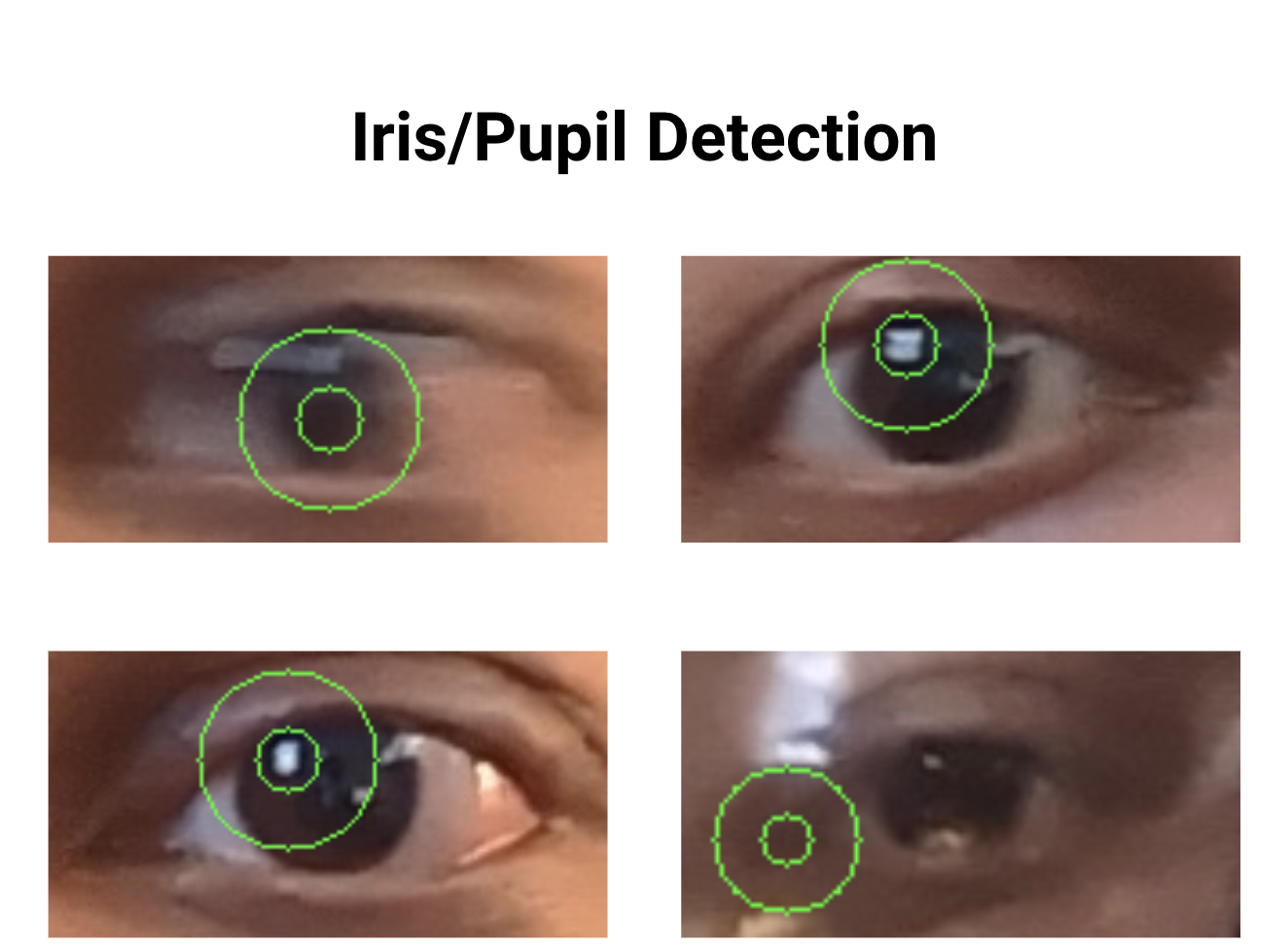}
        \caption{Examples of pictures in which the pupil or iris were incorrectly detected.}
        \label{fig:WrongPred}
    \end{minipage}%
    \hspace{1mm} 
    \begin{minipage}{0.47\textwidth}
        \centering
        \includegraphics[scale=0.285]{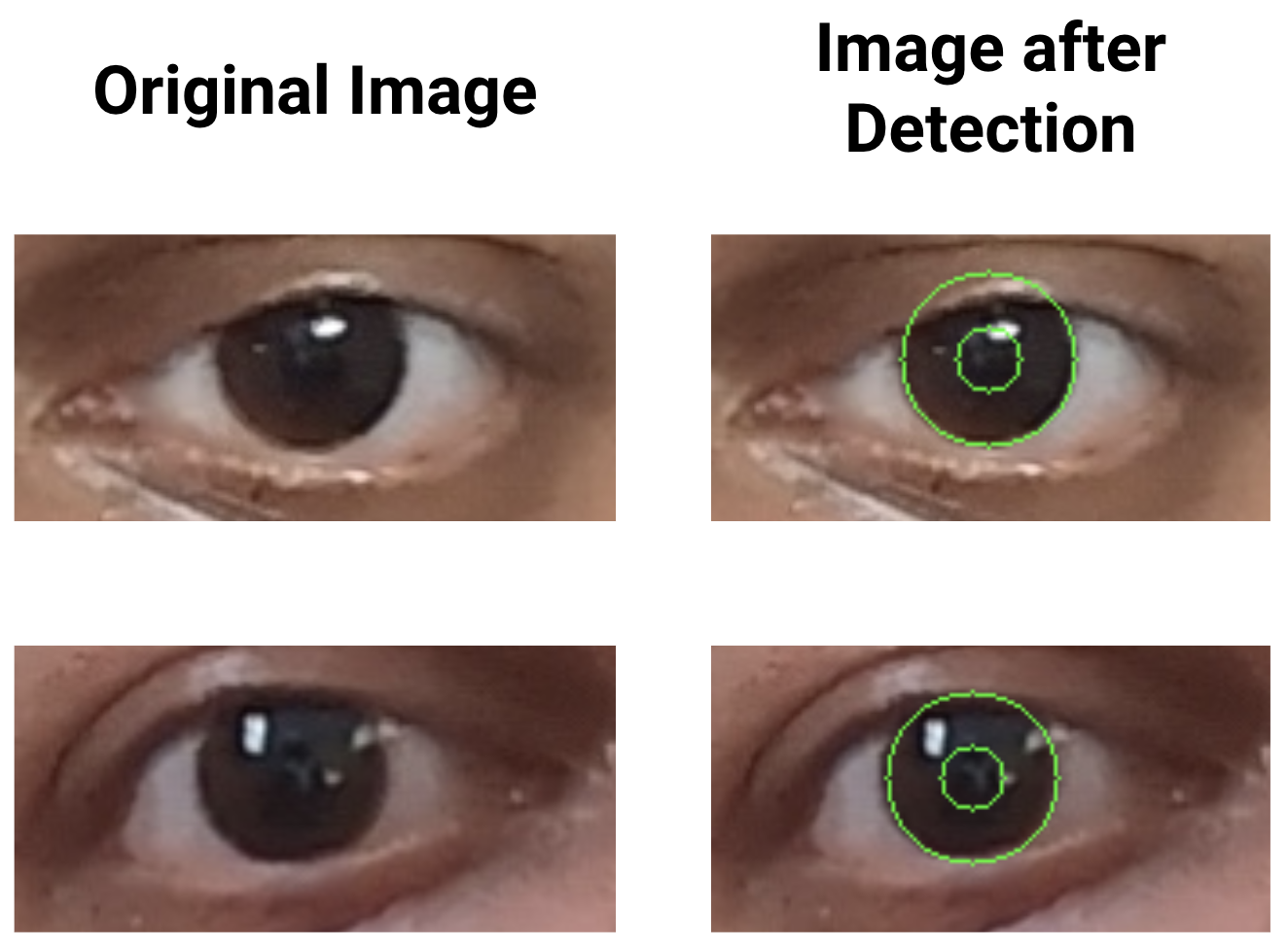}
        \caption{Example of pictures in which the pupil and iris were correctly detected.}
        \label{fig:RightPred}
    \end{minipage}
\end{figure}

To evaluate the performance of the PIR estimation system for obtaining feasibility in real-life deployment. We conducted a second feasibility study with one independent unseen participant (T3: tester 3) not in the first feasibility study. To benchmark the robustness of our method, we collected eye images of the participant in different contexts such as outdoor/indoor, physical activity/sedentary, and low/bright lighting. We ran our PIR estimation pipeline on these and saved the predicted annotated images for later verification. Then, one of the research volunteers manually verifies all the predicted annotations for its accuracy. Figure \ref{fig:WrongPred} shows examples of pictures in which the pupil or iris was incorrectly detected, and Figure \ref{fig:RightPred} shows examples of pictures in which the pupil and iris were correctly detected by our method. Lastly, we benchmark the runtime of our whole pipeline on Google Pixel 4, as shown in Table \ref{tab:PupilSenseRuntimeBenchmark}. Our entire pipeline PIR estimation took 0.7 sec to process one image. We can process multiple images at once by implementing data processing with modern tools such as Apache Airflow.

\begin{table}[h]
\centering
\caption{Runtime Benchmark on Google Pixel 4 and NVIDIA Tesla V100}
\begin{tabular}{lcc}
\toprule
\textbf{Component} & \textbf{Runtime} & \textbf{Avg. Runtime(sec)} \\
\midrule
Image Acquisition & Mobile & 0.4 \\
Eye Region Extraction & Mobile & 0.1 \\
PIR Estimation & Server & 0.2 \\
Full Pipeline & - & 0.7 \\
\bottomrule
\end{tabular}
\label{tab:PupilSenseRuntimeBenchmark}
\end{table}

\section{Depression Study in the Wild}
\subsection{Study}
\subsubsection{Protocol}
The Institutional Review Board (IRB) at the University reviewed and approved the study. Participants in this study were on-boarded remotely across multiple time zones via Zoom conference meetings during the COVID-19 Public Health Emergency (PHE). To be eligible to join, participants had to be 18 or older and have an Android phone with a mobile data plan. The participants were requested to complete a screening survey and choose a suitable time for the onboarding video call. During the onboarding session, the interviewer provided the participants with details of the informed consent and asked them to answer the baseline questions. Following the baseline, the interviewer conducted a semi-structured interview to gain insight into the participant's mental health. Afterward, the study mobile app was installed on the participant's device to collect eye images from their smartphone. The study questionnaires were distributed via notifications on the phone and administered utilizing Qualtrics, an online survey platform.

\subsubsection{Participants}
In total 38 participants were recruited for the study, but only 25 successfully completed it. One participant exited the study after two days at the beginning of the study, citing a significant battery drain caused by our app. However, it was later discovered that this was due to their extensive social media use (one of the triggers that initiated data collection), which frequently activated the data collection mechanism. Personal reasons also led to the withdrawal of 3 participants; 5 failed to complete the study surveys, and 4 had Android versions incompatible with our data collection trigger module, resulting in their exclusion. 

In N=25 participants, the average age was 27.88 years (SD = 8.87), with a range from 18 to 48 years\footnote{9 participants in the demographic survey did not provide age data.}. The gender distribution included 8 females and 11 males, while 6 participants chose not to disclose their gender. Regarding ethnicity, 15 identified as Asian, 4 as Caucasian, and 6 did not specify their ethnicities. Educational background varied: 1 participant had completed high school, 8 held bachelor's degrees, 10 had master's degrees, and 6 did not reveal their highest educational attainment. Concerning mental health history, 4 participants reported a past diagnosis of a mental disorder, 15 reported no such diagnosis, and 6 did not respond to this question. A comprehensive breakdown of these demographics is presented in Table \ref{tab:demographic}.

\begin{table}[h]
\centering
\caption{\label{tab:demographic} Demographic Distribution}  
\setlength{\tabcolsep}{4pt}  
\begin{tabular}{p{3cm}cccc}
\toprule
\textbf{Attribute} & \textbf{Unspecified} & \textbf{Male} & \textbf{Female} & \textbf{Total} \\
\midrule
\textbf{Gender} & 6 & 11 & 8 & 25 \\
\textbf{Age (Average)} & - & 24.11 & 32.71 & 27.88 \\
\textbf{Ethnicity} &  &  &  &  \\
- Unspecified & 6 & 0 & 0 & 6 \\
- Asian & 0 & 9 & 6 & 15 \\
- Caucasian & 0 & 2 & 2 & 4 \\
\textbf{What is your highest education qualification?} &  &  &  &  \\
- Unspecified & 6 & 0 & 0 & 6 \\
- High School & 0 & 0 & 1 & 1 \\
- Bachelor's Degree & 0 & 5 & 3 & 8 \\
- Master's Degree & 0 & 6 & 4 & 10 \\
\textbf{How often do you experience depressive state?} &   &   &   &   \\
- Unspecified & 6 & 0 & 0 & 6 \\
- Not at all often & 0 & 2 & 1 & 3 \\
- Not so often & 0 & 4 & 4 & 8 \\
- Somewhat often & 0 & 5 & 1 & 6 \\
- Very often & 0 & 0 & 2 & 2 \\
\textbf{Have you been diagnosed with any mental health disorder? } &  &  &  &  \\
- Unspecified & 6 & 0 & 0 & 6 \\
- No & 0 & 9 & 6 & 15 \\
- Yes & 0 & 2 & 2 & 4 \\
\textbf{Do you smoke marijuana?} &  &  &  &  \\
- Unspecified & 6 & 0 & 0 & 6 \\ 
- No & 0 & 9 & 8 & 17 \\ 
- Yes & 0 & 2 & 0 & 2 \\
\bottomrule
\end{tabular}
\end{table}

\begin{table*}[h]
\caption{PIR distribution across participants}
\centering 
\begin{subtable}{.9\textwidth} 
\centering 
\begin{tabular}{cccccccccccccc}

\toprule

PID &  & P08 & P10 & P12 & P13 & P14 & P15 & P16 & P17 & P18 & P19 & P20 & P21 
\\
\midrule
PIRleft & 
\multirow{2}{*}{} min & 0.27 &  0.25 &  0.21 &  0.25 &  0.21 &  0.24 &  0.23 &  0.28 &  0.24 &  0.24 &  0.25 &  0.26 \\
& max &  0.50 &  0.47 &  0.60 &  0.53 &  0.55 &  0.38 &  0.50 &  0.36 &  0.45 &  0.38 &  0.37 &  0.43 \\
& mean &  0.35 &  0.32 &  0.34 &  0.33 &  0.33 &  0.32 &  0.33 &  0.32 &  0.33 &  0.31 &  0.33 &  0.34 \\
& std &  0.03 &  0.04 &  0.06 &  0.03 &  0.04 &  0.03 &  0.03 &  0.02 &  0.03 &  0.03 &  0.03 &  0.03 \\
\midrule
PIRright & 
\multirow{2}{*}{} min &  0.23 &  0.29 &  0.21 &  0.27 &  0.20 &  0.23 &  0.20 &  0.29 &  0.22 &  0.28 &  0.24 &  0.24 \\ 
& max  &  0.67 &  0.39 &  0.57 &  0.42 &  0.53 &  0.45 &  0.50 &  0.38 &  0.52 &  0.62 &  0.36 &  0.39 \\
& mean &  0.34 &  0.33 &  0.34 &  0.32 &  0.32 &  0.32 &  0.33 &  0.32 &  0.33 &  0.41 &  0.31 &  0.32 \\
& std &  0.04 &  0.02 &  0.06 &  0.03 &  0.04 &  0.03 &  0.03 &  0.02 &  0.03 &  0.07 &  0.03 &  0.03 \\
\bottomrule
\end{tabular}
\label{subtable1}
\end{subtable}

\vspace{0.5cm} 

\begin{subtable}{\textwidth} 
\centering 
\begin{tabular}{ccccccccccccccccc}

\toprule
PID & & P23 & P24 & P25 & P27 & P28 & P29 & P30 & P31 & P33 & P34 & P35 & P36 & P38 \\
\midrule
PIRleft & 
\multirow{2}{*}{}min&  0.21 &  0.22 &  0.24 &  0.22 &  0.27 &  0.21 &  0.23 &  0.22 &  0.23 &  0.28 &  0.24 &  0.23 &  0.24 \\
& max &  0.58 &  0.46 &  0.39 &  0.52 &  0.40 &  0.58 &  0.46 &  0.51 &  0.53 &  0.45 &  0.60 &  0.52 &  0.39 \\
& mean &  0.34 &  0.31 &  0.31 &  0.33 &  0.34 &  0.33 &  0.32 &  0.34 &  0.33 &  0.34 &  0.34 &  0.34 &  0.31 \\
& std &  0.03 &  0.03 &  0.03 &  0.03 &  0.02 &  0.05 &  0.03 &  0.06 &  0.03 &  0.03 &  0.04 &  0.03 &  0.03 \\
\midrule
PIRright & 
\multirow{2}{*}{}min &  0.21 &  0.22 &  0.24 &  0.23 &  0.24 &  0.22 &  0.20 &  0.21 &  0.24 &  0.27 &  0.25 &  0.21 &  0.23 \\ 
& max &  0.54 &  0.53 &  0.37 &  0.58 &  0.41 &  0.46 &  0.53 &  0.50 &  0.54 &  0.66 &  0.44 &  0.52 &  0.33 \\
& mean &  0.33 &  0.31 &  0.31 &  0.32 &  0.34 &  0.33 &  0.33 &  0.33 &  0.34 &  0.34 &  0.33 &  0.34 &  0.29\\
& std &  0.04 &  0.03 &  0.03 &  0.04 &  0.02 &  0.04 &  0.03 &  0.07 &  0.03 &  0.05 &  0.03 &  0.03 &  0.03 \\
\bottomrule
\end{tabular}
\label{subtable2}
\end{subtable}
\label{tabs:PIRDistribution}
\end{table*}

\subsection{Dataset}
\subsubsection{Depressive Episode (PHQ-9)}
Participants' depressive symptoms were evaluated using a self-administered 9-item Patient Health Questionnaire (PHQ-9) at three key points: the beginning of the study (baseline), two weeks into the study (mid-point), and at the study's conclusion (end-point). The PHQ-9 items are rated on a scale from 0 to 4. Monitoring an individual's PHQ-9 scores over time can offer significant insights into the evolution of their mental health. The PHQ-9 categorizes depression severity into five levels: scores of 0–4 suggest no depression symptoms, 5–9 indicate mild symptoms, 10–14 suggest moderate symptoms, 15–19 are indicative of moderately severe symptoms, and a score between 20–27 reflects severe depressive symptoms.

A depressive episode is defined as a two-week period marked by consistent feelings of sadness, hopelessness, and diminished interest or pleasure in most activities, as indicated by a participant's PHQ-9 score. For our study, we categorized two weeks of a participant's data as either a depressive or non-depressive episode based on their PHQ-9 scores at the start and end of this period. A depressive episode is identified only if the participant's PHQ-9 score is 5 or higher at both the beginning and conclusion of the two-week period. If not, the period is classified as non-depressive. In our analysis, we identified 14 instances of depressive episodes and 30 instances of non-depressive episodes, with each episode spanning a duration of two weeks. With 6 participants dropping after 2 weeks in the study.

\subsubsection{Objective Physiological Measurement}
The participants carried our study app installed on their phones into their daily lives while our app collected their eye images in the background. The participants were not limited, they encountered various environmental conditions such as indoor/outdoor, physical activity/sedentary, and data were collected throughout the day and night. Our study app is designed to collect eye data from the user under two specific conditions: firstly, when the user unlocks their phone, and secondly, when the user opens any of a predetermined set of trigger apps. Upon unlocking the phone, our app becomes active for a duration of 10 seconds. Similarly, it also activates for 10 seconds when the user launches any of the thirty-five selected apps, categorized into groups such as communication, social media, productivity, entertainment, and health. The eye data collection pipeline on phone detect the eyes in the captured frame and sends the cropped eye images for left and right to our research server over a secure line for PIR estimation. Then, these images in one data sample session of 10 seconds are passed through the system described in Section \ref{sec:PIREstimation} for estimating the pupil-iris ratio for both eyes.

\subsection{Data Processing}
Our app collected eye images of the participants for 15995 instances across participants with an average of 639.8 unique interactions with the phone per participant when we collected data. However, our PIR estimation pipeline was only able to infer PIR in 8299 instances of it successfully. On average we computed 11.85 instances of PIR estimation per day per participant. Previous research has shown that the pupil-to-iris ratio generally varies between 0.2 (highly constricted pupil) and 0.7 (highly dilated pupil) \cite{hollingsworth2009pupil, tomeo2015biomechanical, khan2023deformirisnet}. We take all the instances with a pupil-to-iris ratio between 0.2 and 0.7 to create our final dataset with depressive episode ground truth labeling; the PIR distribution per participant is shown in Table \ref{tabs:PIRDistribution}. 

 After applying these inclusion criteria, we identified 6657 instances of PIR estimation in our dataset for further processing.  We then segmented the participants' days into 4 epochs (midnight, morning, afternoon, and evening) at 6-hour intervals. For each epoch, we computed statistical features such as sum, min, max, mean, median, and std for all PIR estimation instances. Due to participant inactivity or failures in the PIR estimation pipeline, data for some participants were not available in all epochs. To address this, we imputed missing values with the mean of the day. In total, we had 48 features for each participant per day, which were then labeled with the depressive episode ground truth. In total, we labeled 528 days from 25 participants. We observed some variations in participation duration, with an average of 3.36 days missing due to participants' early exit from the study. Consequently, we had a total dataset of 616 participant days. Notably, 55 days lacked recorded eye images due to reasons provided by participants, such as planned holidays or breaks, while an additional 33 days lacked data due to factors like image quality and eye-open probability in the PIR extimation pipeline. Therefore, our effective data points for analysis amounted to 528.

\subsection{Statistical Analysis}
The dataset contains PIR features across various epochs of the day. Our primary target variable for statistical analysis was the presence or absence of a depressive episode. Pearson's correlation coefficient (r-value) was calculated against the target variable for each epoch PIR feature in the dataset. Mean and standard deviation were  separately computed for the two groups. Features were then ranked based on the absolute strength of their r-values to identify the top significant features ($p < 0.05$) most strongly correlated with depressive episodes, irrespective of the correlation's direction. We retained only features with at least a weak correlation ($r-value \geq abs(0.20)$) for further analysis and modeling.

\subsection{Feature Selection}
We employed feature selection (FS) using a Random Forest classifier to assess the significance of all features in the dataset and generate a separate set of selected features. Subsequently, We established a threshold based on the mean value of these importance scores (Gini importance). Features with importance values exceeding this mean threshold were considered crucial and retained for further analysis, while those below the threshold were discarded. This approach ensures a data-driven, unbiased criterion for feature selection, improving model interpretability and performance by focusing on features that contribute to predicting depressive episodes.

\subsection{Classification Framework}
Our classification framework uses tree based supervised machine learning (ML) algorithms, specifically Decision Trees, chosen for their high performance demonstrated in previous studies \cite{yang2016decision, islam2018depression} predicting depression. To address the imbalance between depressive and non-depressive episode samples in our dadtaset and mitigate bias, we employed oversampling of the minority class using the synthetic minority over-sampling technique (SMOTE) \cite{chawla2002smote} for the training set. Additionally, we performed hyperparameter tuning to maximize the area under the receiver operating characteristic curve (AUROC). We assessed the predictive performance of our ML models using metrics such as Accuracy, Precision, Recall, F1 score, and AUROC. Finally, we adopted a uniform method where a single classifier for depressive episodes was developed for all users, employing a leave-one-participant-out (LOPO) cross-validation technique, also known as leave-one-out or leave-one-group-out cross-validation. This method, commonly utilized in many mobile inference systems, offers clear insights into model generalizability. Once implemented, this classifier remains consistent without further alterations.

\section{PupilSense Evaluation}

\subsection{Statistical difference in pupillary response between depressive and non-depressive episodes}
\begin{table}[h]
\centering
\caption{\label{tab:corr}Summary of Feature Correlations with Depressive Episodes}  
\setlength{\tabcolsep}{4pt}  
\begin{tabular}{lp{1cm}rp{1.5cm}p{1.5cm}}
\toprule
Feature & $p$-value ($\displaystyle \leq 0.05$) & $r$-value & Depressive Episode Mean (SD) & Non-Depressive Episode Mean (SD) \\
\midrule
    pirRightstd\_morning &                   0.00 &     0.34 &                  0.06 (0.04) &                      0.03 (0.02) \\
   pirRightmean\_morning &                   0.00 &     0.27 &                  0.36 (0.07) &                      0.33 (0.04) \\
 pirRightmedian\_morning &                   0.00 &     0.27 &                  0.36 (0.06) &                      0.33 (0.04) \\
    pirRightmax\_morning &                   0.00 &     0.25 &                    0.4 (0.1) &                      0.36 (0.06) \\
    pirLeftmean\_morning &                   0.00 &    -0.25 &                  0.32 (0.04) &                      0.34 (0.03) \\
  pirLeftmedian\_morning &                   0.00 &    -0.24 &                  0.32 (0.03) &                      0.34 (0.03) \\
        pirRightsum\_evening &                   0.00 &     0.24 &                  2.76 (2.54) &                      1.72 (1.67) \\
         pirLeftsum\_evening &                   0.00 &     0.23 &                  2.76 (2.52) &                      1.73 (1.66) \\
     pirLeftmin\_morning &                   0.01 &    -0.21 &                   0.3 (0.05) &                      0.31 (0.04) \\
\bottomrule
\end{tabular}
\end{table}

\textcolor{black}{The analysis explored the relationship between various features and depressive episodes, as shown in Table \ref{tab:corr}. These features were sorted based on the magnitude of their correlation (r-value) with the target variable, which signifies the occurrence of a depressive episode. We identify features with at least a weak correlation ($r-value \geq abs(0.20)$) in correlation analysis to prioritize meaningful relationships and enhance model interpretability and efficiency while reducing noise and the risk of overfitting. Our correlation analysis yielded mixed results, indicating a positive relationship (more pronounced pupil dilation) between depressive episodes and right eye PIR, and a negative relationship (less pronounced pupil dilation) with left eye PIR observed during morning hours (6 am - 12 pm). These findings are consistent with previous clinical studies, which have reported varied outcomes. For example, one study \cite{schneider2020pupil} found less pronounced pupil dilation in anticipation of reward in acutely depressed patients, while others \cite{burkhouse2015pupillary} have observed increased pupil dilation to sad stimuli in children at high risk for depression. The presence of different features across various times of the day (morning and evening) suggests that the impact of depression on pupillary response may fluctuate throughout the day, potentially reflecting circadian rhythms or fluctuating levels of fatigue and alertness \cite{tseng2018alertnessscanner}. Overall, the analysis indicates a significant relationship between pupillary response and depressive episodes, with variations in the strength and direction of this relationship across different times of the day and different pupillary response metrics. This finding aligns with and contributes to the expanding body of research on the physiological manifestations of depression.}

\subsection{Detecting Depressive Episode in the Wild}
\textcolor{black}{When evaluating the performance (refer to Table \ref{tab:results_features}) of our models for detecting depressive episodes, we observe that the Top Significant Features (TSF) model outperforms others across all metrics. We achieved the highest accuracy with TSF, reaching 76\%, indicating its superior reliability in correctly classifying depressive and non-depressive instances. With the highest precision and recall, at 64\% and 63\% respectively, the TSF model demonstrates a balanced approach, effectively minimizing false positives and false negatives. This balance is critical in interventions, ensuring  that the majority of detected depressive episodes are true positives while also capturing a high proportion of actual cases. In practice, the TSF model's high F1 score (0.64) and AUROC (0.71) signify robustness, making it the preferred choice when misdiagnosis consequences are significant, such as in mental health resource allocation or initiating treatment protocols. Deploying this model could enhance patient outcomes, increasing the likelihood of timely and appropriate interventions to those in need without overburdening healthcare systems with false positives.}


\begin{table}[h]
\caption{\label{tab:results_features} Model Performance in our Depressive Episode Detection} 
\centering
\footnotesize
  
    \begin{tabular}{p{3.5cm}p{0.5cm}p{0.5cm}p{0.5cm}p{0.5cm}p{0.8cm}p{0.8cm}}
    \toprule
    Feature Set & Acc. & Prec. & Rec. & F1 & AUROC \\ 
    \midrule
    
    Feature Selection (FS) & 0.68\ & 0.53\ & 0.57\ & 0.55\ & 0.62\ \\ 
    Top Significant Features (TSF) & \textbf{0.76}\ & \textbf{0.64}\ & \textbf{0.63}\ & \textbf{0.64}\ & \textbf{0.71}\ \\ 
    All & 0.67\ & 0.50\ & 0.39\ & 0.44\ & 0.51\  \\ 
    
    \bottomrule
 
\end{tabular}
\end{table}

\section{Discussion}
In this section, we present our study on detecting depressive episodes using pupillary response measurement in real-world settings. Our contributions are manifold: we first developed an automated PIR estimation system in a naturalistic environment, and applied this system for depression modeling in real-world contexts. Our correlation analysis unveiled that variations in pupillary response at different times of the day (morning and evening) can distinguish between individuals with depressive and non-depressive states. This suggests that the impact of depression on pupillary response may fluctuate throughout the day, potentially reflecting changes in circadian rhythms or varying  levels of fatigue and alertness \cite{tseng2018alertnessscanner}. Noticeably, we observed a negative correlation between PIR from the left eye and a positive correlation between PIR from the right eye. However, we were unable to confirm the significance of this disparity based solely on the available ground truth and baseline participant information, indicating a need for further investigation. Finally, we developed three distinct models. The top-performing model, leveraging the most significant features, achieved an AUROC of 0.71. Our model outperformed one constructed solely with sensor data (using AWARE \cite{ferreira2015aware}) from Bluetooth (Accuracy=69.3, F1=0.64), Calls (Accuracy=68.5, F1=0.59), GPS (Accuracy=69.5, F1=0.62), and Steps Counter (Accuracy=63.6, F1=0.53), achieving a higher F1 score of 0.64, and an  accuracy of 76\%, as reported in prior research by Chikersal et al. \cite{chikersal2021detecting}. However, in this study where all sensors were combined, a better F1 score (0.78) was reported compared to our model. Furthermore, our system outperformed a recent comparable system that uses facial images captured by a smartphone's front camera to detect depression \cite{nepal2024moodcapture}, achieving only a balanced accuracy of 60\%. In another recent study \cite{price2024detecting}, the best model for detecting depression using passively-collected wearable movement data received an F1 score of 0.68. This suggests that our pupilometry approach could complement  previous work using smartphone \cite{chikersal2021detecting, nepal2024moodcapture} and wearable \cite{pedrelli2020monitoring, price2024detecting} data sources such as GPS, App Usage, Steps, Bluetooth, skin conductance, and EDA, thereby demonstrating its incremental utility.

Previous studies have utilized various methods to assess depression, including questionnaires, social, behavioral, and physiological signals. Recently, researchers have proposed methods to predict depressive states in real-world settings by passively collecting social and behavioral data from smartphones. However, current mobile sensing methods \cite{opoku2022mood, chikersal2021detecting, farhan2016behavior, nepal2024moodcapture} encounter challenges in accurately detecting physiological signals \cite{abdullah2018sensing}, particularly those related to involuntary pupillary responses, which are significant markers of depression. While wearable sensors have been proposed for physiological monitoring, the substantial costs associated with purchasing (\$1,690 per device) and using these devices pose a major obstacle, as used in prior studies \cite{pedrelli2020monitoring}. On the other hand, pupillary response-based alertness prediction has shown significant promise in capturing physiological signals \cite{tseng2018alertnessscanner}. Nevertheless, the practical implementation of these studies in real-world settings is limited due to factors such as computational expenses, device costs, and the additional effort required by users to wear extra equipment. Therefore, our research aims to address these shortcomings by introducing and evaluating the practicality of using pupillary response data to identify depressive episodes in real-world, naturalistic settings.

\subsection{Ethical Considerations and Privacy}
During the development of our system, we placed a high priority on ethical considerations, with a particular focus on privacy and data security. To this end, participants received a comprehensive briefing about the objectives of the study, the type of data that would be collected, and how it would be used, thereby guaranteeing that consent was fully informed. Our system captures facial images (later removed) to extract and crop eye images for subsequent processing. Such data acquisition methods can raise numerous concerns about individual privacy, especially when dealing with stigmatized topics like mental health within the research community. However, we do not store identifiers that directly link individuals to eye images. Researchers in the Human-Computer Interaction (HCI) field should prioritize creating user nudges to strike a balance between privacy and data quality in modeling, as advocated in studies on privacy-preserving systems \cite{denning2014situ, balebako2014improving, felt2012ve}. These nudges would empower users to actively manage and be informed about system operations, particularly in private settings where discomfort may arise, even in cases where no images are being collected or stored. This approach entails providing users with clear information regarding the status of data collection. Furthermore, recognizing the importance of users' privacy perception is crucial, especially in finance-related apps or others where data collection sensitivity is pronounced. The risk that data collection could inadvertently reveal financial statuses or personal conditions highlights the urgent need for mechanisms that empower users to actively manage data collection. To mitigate these concerns, it is essential to offer users the ability to disable tracking or data collection at specific times or within certain applications. A 'Do Not Track' feature represents an ideal solution, allowing users to opt-out of data collection during sensitive periods or in particular apps. Such a feature, with its adaptability to specific contexts and user-defined settings, acknowledges the dynamic nature of user consent and preferences. By enabling a more nuanced control over data privacy, this approach not only addresses the intricate nature of privacy concerns but also actively empowers users, safeguarding them against potential personal implications of data collection.

While recognizing the financial aspects, it is also imperative to carefully consider the potential privacy ramifications of camera-based sensing systems that continuously collect user data throughout the day. The privacy concern lies in the possible risk of unauthorized access or misuse of personal and sensitive visual data captured by these cameras. To address such challenges, one strategy involves conducting all necessary feature extraction directly on the user's device, ensuring that only the user's device handles identifiable images. Additionally, once image processing is completed, these images should be immediately deleted from the device's memory. In the future work section, we discuss solutions to overcome these obstacles, aiming to advance non-invasive physiological depression sensing in a manner that safeguards privacy. Enhancing privacy in such systems could involve adopting frameworks that consolidate data collection through cluster heads within sensor networks. This approach facilitates efficient and anonymous data aggregation. In addition, applying differential privacy to Eigenfaces in facial recognition can significantly improve the security of biometric data, protecting against unauthorized access and attacks \cite{usman2019p2dca, chamikara2020privacy}. These strategies, as discussed in our future work, pave the way for more secure privacy-preserving systems .

\section{Limitations and Future Work}
As our results demonstrate the feasibility of implementing a depressive episode classification system on readily available smartphones combined with a GPU server instance, the initial implementation's insights and outcomes will guide further development and the launch of the PupilSense System. The PupilSense prototype is implemented on the Android API 21 and utilizes Detectron2 running on NVIDIA Tesla V100. PIR estimation and model training are conducted offline, and future work will explore online training and a quantized model for PIR estimation using a mobile Detectron2 model like D2GO \cite{d2go} that runs on the user's device, thereby mitigating potential user privacy concerns arising from eye images being transferred to an external server for processing. In addition, future work will also consider increasing the frequency of image acquisition (currently at 2.5 Hz) and processing through codec, enabling researchers to capture more physiological signals from the eye, such as blink rate \cite{jongkees2016spontaneous, al2018depression}, eye-aspect ratio \cite{mehta2019real, maior2020real}, percentage of eyelid closure over the pupil over time (PERCLOS) \cite{chang2022drowsiness, sommer2010evaluation} etc., which have been found to be effective in detecting cognitive function, drowsiness, and fatigue.

Our approach aligns with the literature suggesting that naturalistic observations can provide unique insights into the complexities of human behavior and physiological responses. As noted by Picard et al. \cite{picard2016automating}, everyday environments could reveal behavioral and physiological response patterns unobservable in controlled settings, such as stress cases. Additionally, the dynamic nature of the pupillary response suggests its sensitivity to a wide variety of stimuli, including those typically encountered while using a smartphone \cite{miranda2018eye}, reading a book and using a smartphone \cite{mirzajani2022dynamic}, and modeling user preferences during smartphone browsing \cite{shen2022pupilrec}. While we acknowledge that the lack of controlled stimuli presents challenges in directly comparing our findings with prior work, it also presents a unique opportunity to investigate the effect of real-world interactions on physiological indicators of depression. Our study is in line with research in passive sensing \cite{pedrelli2020monitoring, chikersal2021detecting, renn2018smartphone}, supporting the idea that understanding physiological responses to everyday, uncontrolled stimuli is crucial for developing more effective interventions for depression in real-life settings. Furthermore, we propose future research to explore controlled stimuli in naturalistic settings, aiming to bridge the gap between laboratory-based and real-world observations.

Although PupilSense accounts for ambient light and various activities, further improvements are needed for handling extremely dark or bright light conditions. Image quality degradation in dark environments remains a challenge that could be mitigated by refining the image capture process with additional infrared light source. The proliferation of commercial smartphones featuring infrared emitters and cameras for identity verification could help overcome challenges associated with extreme lighting conditions. Moreover, it is important to recognize that pupil size variations extend beyond depressive symptoms, encompassing factors such as age \cite{winn1994factors} and mind wandering \cite{bruineberg2021habitual} during smartphone use, among other factors. Given our study's participants fell within a narrow age range (average: 27.88 years), we advocate for data collection from a more diverse population spanning various age groups to minimize age-related influences. To differentiate mind wandering from depressive symptoms effectively, future research should integrate additional contextual cues. This may involve considering factors such as blink rate \cite{jongkees2016spontaneous, al2018depression}, eye-aspect ratio \cite{mehta2019real, maior2020real}, PERCLOS \cite{chang2022drowsiness, sommer2010evaluation}, and patterns of app usage \cite{gordon2019app}. Incorporating these factors could augment the efficacy of the proposed method.
\section{Conclusion}
Depressive episodes serve as critical indicators in mental health disorders such as Major Depressive Disorder and Bipolar Disorder, where early detection is essential for effective intervention. Current methods for monitoring these episodes often involve wearable devices or self-report questionnaires, such as the PHQ-9. While valuable, these techniques can be intrusive or heavily reliant on subjective self-reporting, potentially failing to accurately capture subtle, ongoing changes in an individual's mental state. This limitation underscores the urgent need for more discreet and continuous monitoring methods. To address this need, our research introduced PupilSense, a mobile system seamlessly integrated into daily life. By utilizing real-time pupillary response tracking through day-to-day smartphone interaction, PupilSense offers a less intrusive and more consistent approach to monitoring depressive episodes (AUROC = 0.71) in naturalistic settings.

\section*{Acknowledgment}

We sincerely thank our research volunteers, Shahnaj Laila, Priyanshu Singh Bisen, and Ishan Garg for their kind assistance in conducting this study. We are also grateful to the participants who generously agreed to share their photos for this study.

\bibliographystyle{IEEEtran}
\bibliography{ref}

\end{document}